# Inverse-designed flat lens for imaging in the visible & near-infrared with diameter > 3mm and NA=0.3

*Monjurul Meem, [1] Sourangsu Banerji, [1] Apratim Majumder, [1] Christian Pies,[2] Timo Oberbiermann,[2] Berardi Sensale-Rodriguez, [1] and Rajesh Menon, [1, 3 a)]*

[1]Department of Electrical and Computer Engineering, University of Utah, Salt Lake City, UT 84112, USA.

[2]Heidelberg Instruments Mikrotechnik, Heidelberg, Germany.

[3]Oblate Optics, Inc. San Diego CA, USA.

[a)] rmenon@eng.utah.edu

**ABSTRACT**

It is generally thought that correcting chromatic aberrations in imaging requires multiple surfaces. Here, we show that by allowing the phase in the image plane of a flat lens to be a free parameter, it is possible to correct chromatic aberrations over a large continuous bandwidth with a single diffractive surface. We experimentally demonstrate imaging using a single flat lens of diameter > 3mm, focal length = 5mm (NA = 0.3, f/1.59) that is achromatic from 0.45μm to 1μm. This combination of size, NA and operating bandwidth has not been demonstrated in a flat lens before. We experimentally characterized the point-spread functions, off-axis aberrations and the broadband imaging performance. In contrast to conventional lens design, we utilize inverse design, where phase in the focal plane is treated as a free parameter. This approach attains a phase-only (lossless) pupil function, which can be implemented as a multi-level diffractive flat lens that achieves achromatic focusing and imaging.

**INTRODUCTION**

The lens is considered the most fundamental element for imaging. Imaging is information transfer from the object to the image planes. This can be accomplished via a conventional lens that essentially performs a one-to-one mapping [1], via an unconventional lens (such as one with a structured point-spread function or PSF) that performs a one-to-many mapping, or via no lens, where the light propagation essentially performs a one-to-all mapping. In the first case, the image is formed directly. In the second case, the image is formed after a computation and can be useful, when encoding spectral [2-3] or depth [4] or polarization [5] or other information into the geometry of the PSF itself. Note that the modification of the PSF may be at the same scale as the diffraction limit [2-5] or it can even be much larger [6-9]. The image can be recovered in many cases in the optics-less scenario as well [10-11], but more importantly, machine learning may be employed to make inferences based on the acquired information (even without performing image reconstruction for human visualization), which has potential implications for privacy among other interesting outcomes [12]. However, the conventional lens approach is preferred in many cases due to the high signal-to-noise ratio achievable at each image pixel (resulting from the 1:1 mapping). When this conventional lens is illuminated by a plane wave, it forms a focused spot at a distance equal to its focal length.

Now, if we appeal to the fact that in the vast majority of imaging applications, only the intensity is measured, the spatial distribution of phase[1] in the focal plane can be an arbitrary function. Then, it is easy to see, for instance via the inverse diffraction transform, [13] that the spatial-phase distribution of the plane wave after it transmits the lens (henceforth, we refer to this as the pupil function) can have multiple forms. In conventional optics, the positive lens is a device that converts incident plane waves into converging spherical waves, whose pupil function therefore, follows a hyperbolic relationship to the radial co-ordinate (see black lines in insets of Fig. 1). However, our argument above shows that this picture is incomplete and the set of pupil functions for an ideal lens is, in fact, infinite. Another way to visualize this is that the converging spherical wave is just one of an infinite set of waves that can converge to a focus, *i.e.*, an intensity distribution that is sharply localized in space. This key insight can be exploited to search for pupil functions that enable achromatic focusing [14-18], extreme depth-of-focus [19] or even high efficiency at high-NA [20]. In this paper, we utilize this concept and inverse design to create a flat lens with diameter=3.145mm, focal length=5mm, NA=0.3 and operating bandwidth of $\lambda$=0.45μm to 1μm (simulated point-spread functions are in Fig. 1). In comparison, the broadest bandwidth visible metalens that we are aware of has a diameter of only 44μm, NA = 0.125 and operating bandwidth of $\lambda$=0.4μm to 0.65μm [21]. Furthermore, metalenses require high-refractive index dielectrics (GaN is used in the example above) that requires expensive semiconductor processing, which makes it impractical for larger-area optics [22]. In contrast, the flat multi-level diffractive lenses (MDLs) described here can be inexpensively replicated into low-index polymers, even over large areas with high precision.

---

[1] Here we refer to the phase of a scalar electromagnetic field, but the argument is equally valid for vector fields as well.

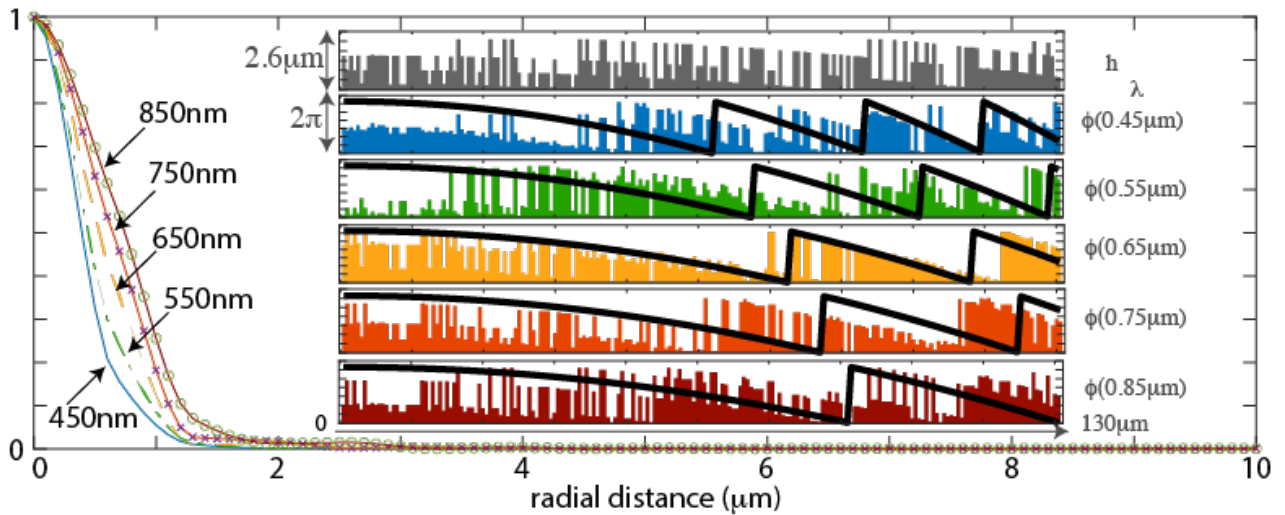

*Figure 1:* *Simulated point-spread functions (PSFs) of MDL for different wavelengths. Insets show the height and corresponding phase distributions for the innermost rings (within a radius*

*of 130μm). See top-view of fabricated device in Fig. 2a. The ideal-lens complex pupil function (hyperbolic function) is plotted for each wavelength with red dashed lines for comparison.*

In Fig. 1, we present the simulated point-spread functions (PSFs) of the MDL for λ = 0.45μm to 0.85μm. In the inset, we also plotted the radial-ring-height distribution and the corresponding phase-shift distributions for λ = 0.45μm to 0.85μm for the inner 201 rings (radius=130μm). The hyperbolic phase function for each wavelength is plotted (solid black lines) for comparison. It is defined as ϕ = -2π/λ*r²/(2f), where is the radial coordinate and f is the focal length, and as expected varies with wavelength, which makes it infeasible to design achromatic diffractive lenses.

The MDL is comprised of 2419 concentric rings of fixed width (0.65μm) and varying heights (0 to 2.6μm). Details of our design procedure has been described elsewhere [22]. The MDL was patterned using high-resolution optical-grayscale lithography in a positive-tone photoresist (maP1200G), whose dispersion was used for design [20]. An optical micrograph of the fabricated device and the measured point-spread functions at λ=0.45μm to 1μm in increments of 50nm are shown in Figs. 2a and 2b, respectively. Achromatic focusing with spot size close to that predicted by the diffraction limit is observed. The illumination was achieved via a collimated beam from a super-continuum source coupled to tunable filters [14].

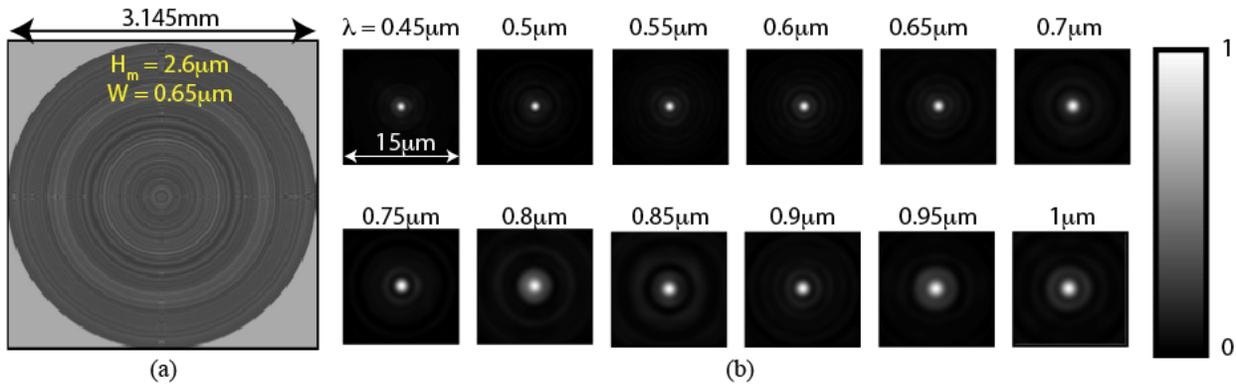

*Figure 2:* Experimental achromatic focusing. (a) Optical micrograph of fabricated lens. Hm is the maximum ring height and W is the minimum ring width. (b) Measured PSFs in the visible-NIR band. The illumination bandwidth of each wavelength was 15nm.

The imaging performance of the MDL was characterized next by measuring the off-axis PSFs under broadband illumination (0.45μm to 0.85μm). The angle of incidence of the plane wave illumination was adjusted from $0^0$ to $30^0$ (Fig. S1) and the corresponding PSFs were recorded on a monochrome image sensor (see Fig. 3a). Off-axis aberrations become apparent at angles >= $15^0$. The modulation-transfer functions (MTFs) were extracted from the measured PSFs (Fig. 3b) and indicate that contrast of 10% is maintained for all angles less than ~$15^0$, from which we can conclude that the estimated full field of view is ~$30^0$. Note that no special effort was made to correct for off-axis aberrations in this design, which if performed properly should significantly increase this value. The linearity of the lens is confirmed by plotting the centroid of the measured PSFs as a function of incident angle (Fig. 3c). Finally, in Figs. 3d and 3e, we show images formed by this lens in the visible and NIR bands of the Macbeth color chart and the Airforce resolution chart, respectively.

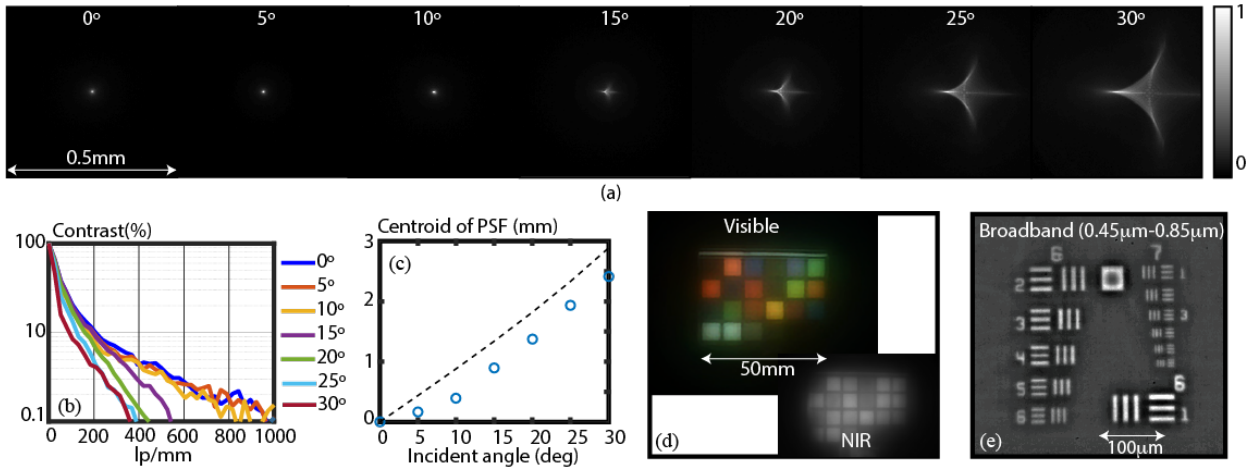

*Figure 3:* Experimental imaging performance. (a) Off-axis PSFs under broadband (0.45μm to 0.85μm) illumination. (b) MTF curves at various angles of incidence corresponding to PSFs in (a). (c) Centroid of the PSF as a function of incident angle demonstrating linearity. (d) Visible

(white LED illumination) and NIR (850nm LED flashlight illumination) images of the Macbeth chart. (e) Broadband image of resolution chart. Experimental details are in Fig. S2.

**Conclusions**

In this Letter, we demonstrated that a single MDL can be made achromatic over a continuous spectrum from 0.45μm to 1μm with diameter>3mm and NA=0.3. Such a lens is made possible via a new approach to design that invokes the non-uniqueness of the lens pupil function. To re-iterate this point, we show similar performance from a second MDL with the same design specifications as the one above, but with a different lens pupil function (Figs. S3, S4). In addition, we have fabricated other visible-NIR MDLs and characterized their aberrations performance (Figs. S6, S7, S8) [14]. All our experimental results confirm that a single diffractive surface is sufficient to correct for aberrations including chromatic aberrations for a large majority of imaging applications over a continuous spectrum.

It is important to note that theoretical work by Stamnes and co-workers identified the "perfect" wave for maximal concentration of power is, in fact, the diverging electric-dipole wave, [23, 24] which are not strictly spherical. In our work, rather than opting for the global maximum of concentration, we design our MDLs to achieve sufficient light concentration at the focus to enable imaging performance. This allows us to incorporate fabrication constraints into the design process, enabling useful devices.

**Acknowledgements**

We thank Brian Baker, Steve Pritchett and Christian Bach for fabrication advice, and Tom Tiwald (Woollam) for measuring dispersion of materials. We would also like to acknowledge the support from Amazon AWS (051241749381). RM and BSR acknowledges funding from the Office of Naval Research grant N66001-10-1-4065 and from an NSF CAREER award: ECCS

#1351389, respectively.

**Competing Interests Statement**

RM is co-founder of Oblate Optics, Inc., which is commercializing technology discussed in this manuscript. The University of Utah has filed for patent protection for technology discussed in this manuscript.

**Materials and Correspondence**

Correspondence and materials requests should be addressed to RM at rmenon@eng.utah.edu.

# Supplementary Information

# Inverse-designed flat lens for imaging in the visible & near-infrared with diameter > 3mm and NA=0.3

*Monjurul Meem, [1] Sourangsu Banerji, [1] Apratim Majumder, [1] Christian Pies,[2] Timo Oberbiermann,[2] Berardi Sensale-Rodriguez, [1] and Rajesh Menon, [1, 3 a)]*

[1]Department of Electrical and Computer Engineering, University of Utah, Salt Lake City, UT 84112, USA.

[2]Heidelberg Instruments Mikrotechnik, Heidelberg, Germany.

[3]Oblate Optics, Inc. San Diego CA, USA.

[a)] rmenon@eng.utah.edu

## 1. Focal spot characterization

The MDL was illuminated with expanded and collimated beam from a SuperK EXTREME EXW-6 source (NKT Photonics). The wavelength and bandwidth was tuned with SuperK VARIA filter (NKT Photonics) for visible wavelengths (400nm-800nm) and SuperK SELECT filter (NKT Photonics) for near infrared wavelengths (800nm-1400nm). The focal spots of the flat lenses were magnified using a 50x objective (LMPLFLN50x, Olympus) and tube lens (ITL200, Thorlabs) and imaged onto monochrome sensor (DMM 27UP031-ML, Imaging Source). The exposure time of the sensor was carefully adjusted to avoid pixel saturation. A dark frame was also captured with all the light source turned off. The gap between objective and tube lens was ~120 mm and that between the sensor and the backside of tube lens was about 180mm. The magnification of the microscope was found by imaging a calibration sample with known feature dimensions. For off axis measurement, the lens was illuminated an angle and the focus spot was directly captured onto the monochrome sensor without magnification. The captured off axis focal spots were up-sampled accordingly to get the modulation transfer function, presented in Fig. 3(b) of the main text.

After capturing the focal spot, the Focusing efficiency was then calculated using the following equation: Focusing efficiency = (sum of pixel values in 3*FWHM) / (sum of pixel values in the entire lens area)

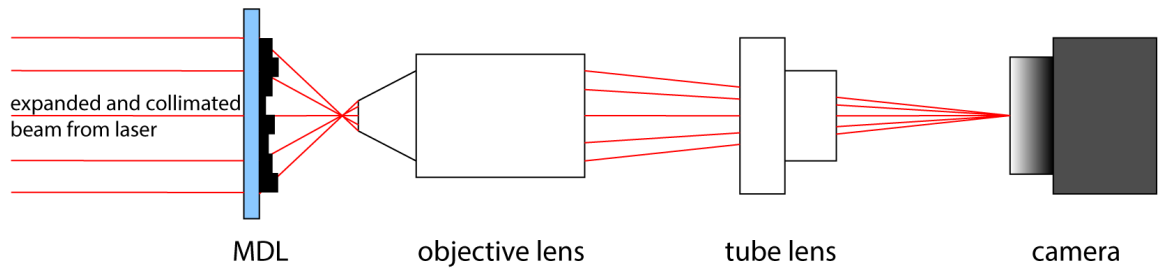

**Figure S1:** Schematic of the set-up used to capture the focal spots of the MDL. An expanded and collimated laser beam is focused by the MDL. An objective lens paired with a tube lens is used to form a magnified image of the focal spot on a CMOS camera.

## 2. Imaging Setup

For imaging, colorful objects, the objects were placed in front of the MDL and were illuminated with white LED flash light for color image and NIR 850nm flash light for infrared images, corresponding images were captured using a color sensor (DFM 72BUC02-ML, Imaging Source).

For resolution chart imaging, the 1951 USAF resolution test chart (R3L3S1N, Thorlabs) was used as object. The USAF target was illuminated with broadband (450nm-850nm) laser and corresponding images were captured using a monochrome sensor (DMM 27UP031-ML, Imaging Source). The exposure time was adjusted to ensure that the images does not get saturated. In each case, a dark frame was recorded and subtracted from the images.

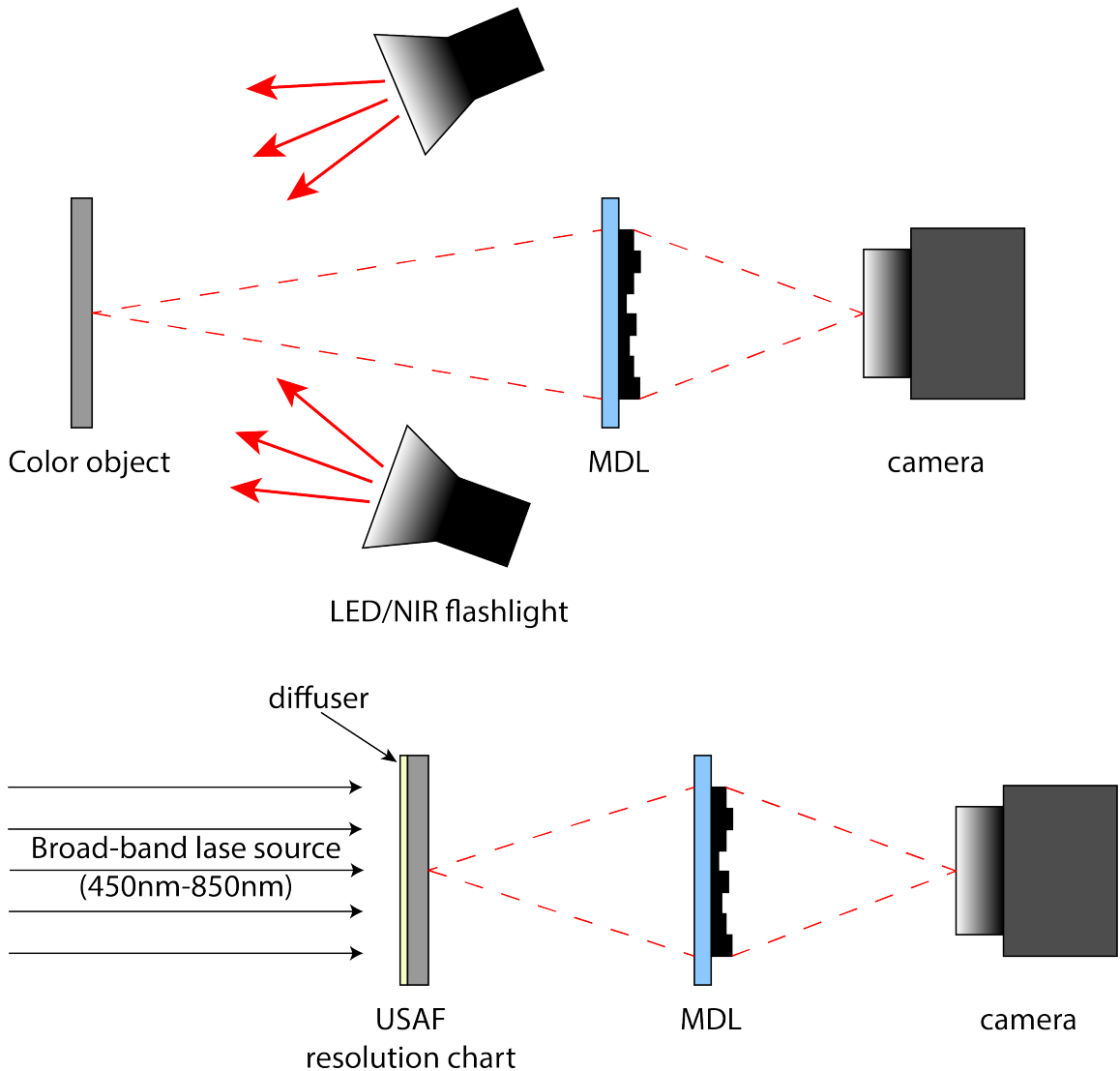

**Figure S2:** Schematic of the imaging setup. The object was illuminated with LED flash light (or NIR flash light) or broad band laser, and the corresponding image formed by the MDL was captured on a camera.

## 3. Performance of Second NA 0.3 MDL

We have designed a second MDL with the same design specification as the one presented in the main text, but with different lens pupil function. The performance of this lens is summarized below.

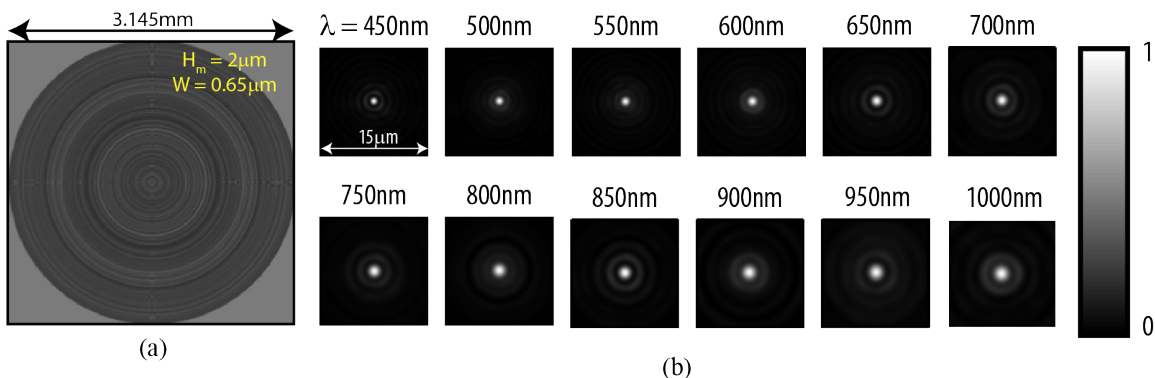

**Figure S3:** Experimental achromatic focusing. (a) Optical micrograph of fabricated lens. Hm is the maximum ring height and W is the minimum ring width. (b) Measured PSFs in the visible-NIR band. The illumination bandwidth of each wavelength was 15nm.

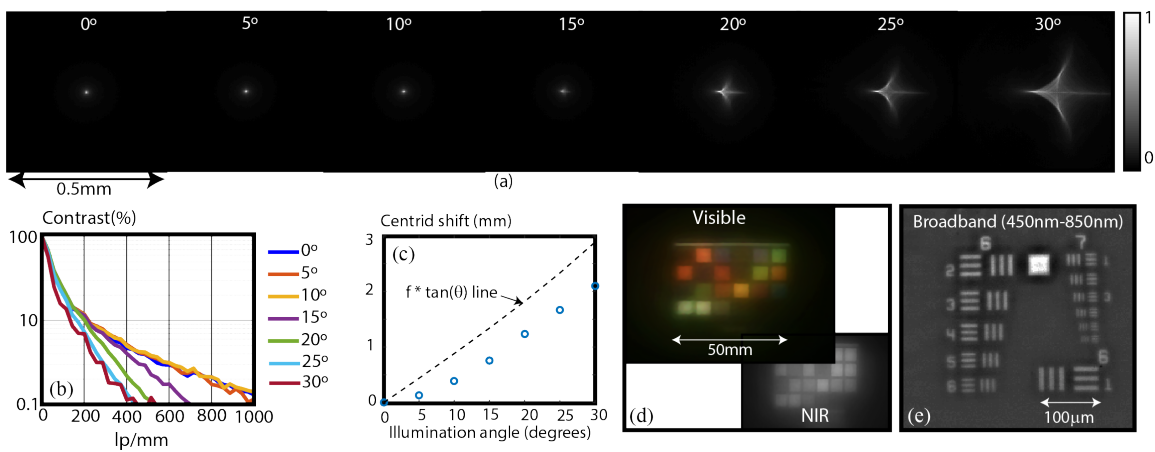

**Figure S4:** Experimental imaging performance. (a) Off-axis PSFs under broadband (0.45μm to 0.85μm) illumination. (b) MTF curves at various angles of incidence corresponding to PSFs in (a). (c) Centroid of the PSF as a function of incident angle demonstrating linearity. (d) Visible (white LED illumination) and NIR (850nm LED flashlight illumination) images of the Macbeth chart. (e) Broadband image of resolution chart.

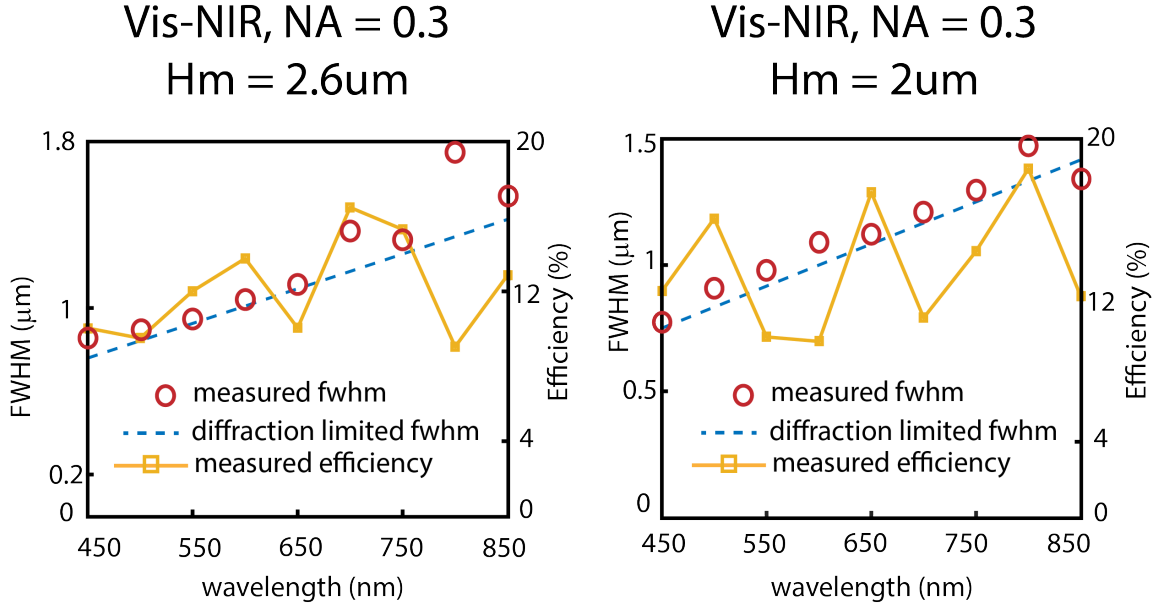

**Figure S5:** Experimental full width half maximum (fwhm) and efficiency characterization. Left one corresponds to the MDL presented in the main text and the right one is for the second MDL.

### 4. Performance of NA 0.075 MDL

We have designed and characterized another broad band MDL with focal length and NA of 1mm and 0.075 respectively. The wavelength of operation is 450nm to 850nm. The performance of this lens is summarized below. Further details can be found in [arxiv paper]

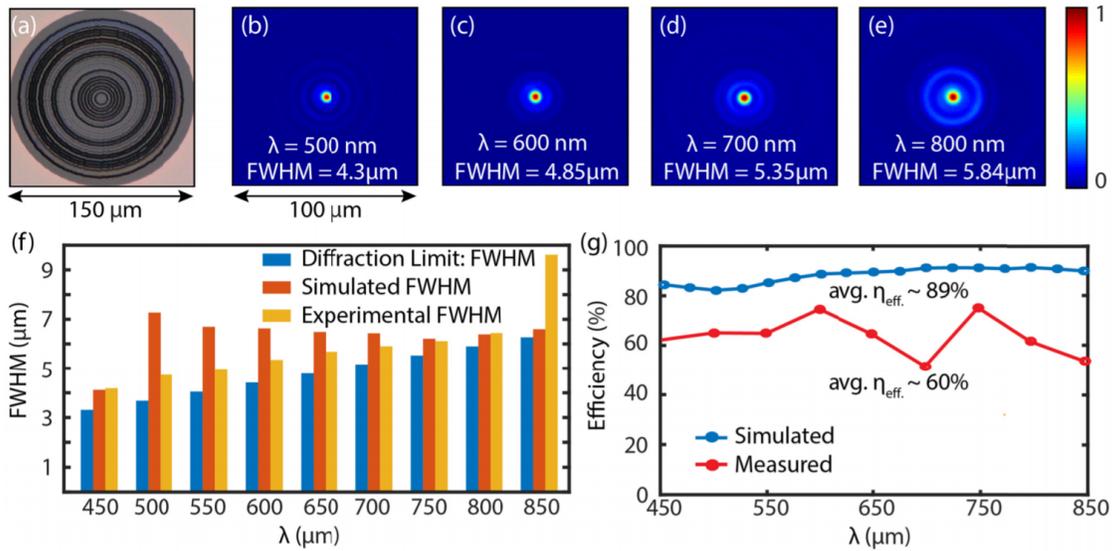

**Figure S6:** Focusing performance of low NA MDL. (a) Optical micrograph of the fabricated MDL. (b-e) Measured point-spread functions at the same focal plane, 1mm away from the MDL at four wavelengths of interest. (f) Measured, simulated and diffraction-limited full-width at half maximum (FWHM) as a function of wavelength. (g) Measured

and simulated focusing efficiency as a function of wavelength. The operating bandwidth of the MDL is 450nm to 850nm.

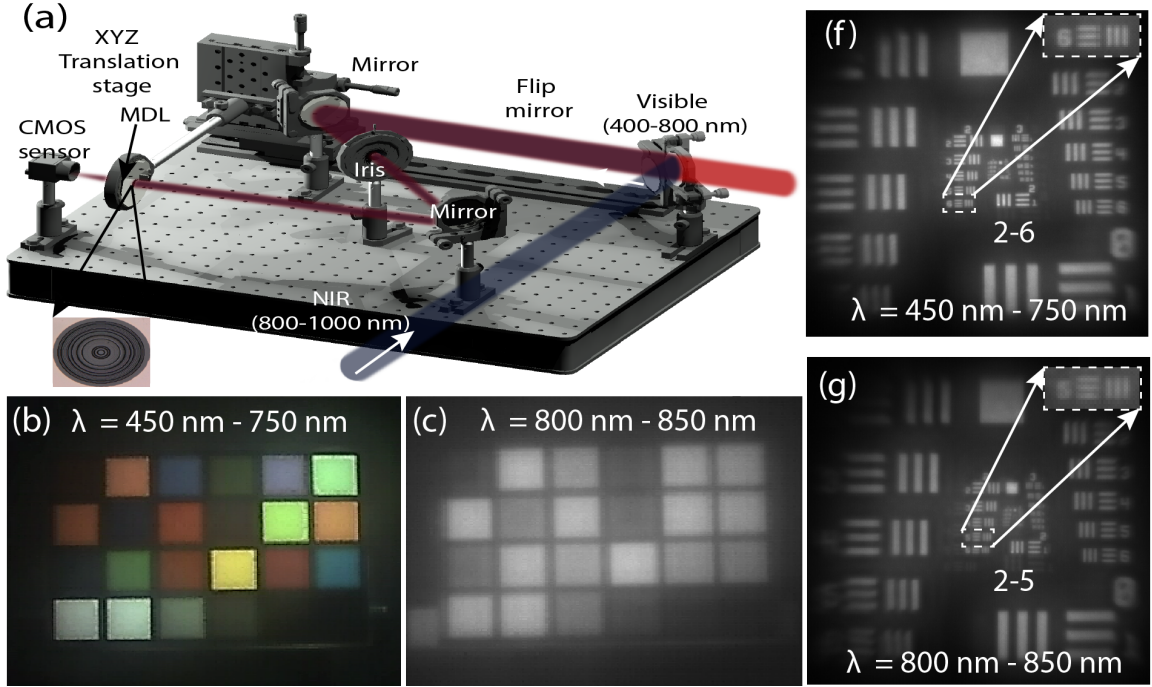

**Figure S7: (a)** Schematic depicting the optical setup used for imaging with the low NA MDL. Exemplary still images for the **(b)** broadband MDL at λ = 450 - 750 nm as well as **(c)** λ = 850 - 850 nm respectively are shown. Resolution of the designed **(f-g)** broadband MDLs under differrent illumination conditions were ascertained by imaging the USAF resolution chart. Insets of **(f-g)** highlight the smallest grating lines (group number-element number) which could be satisfactorily identified with the MDLs. In the all the cases, diffraction limited imaging performance is showcased.

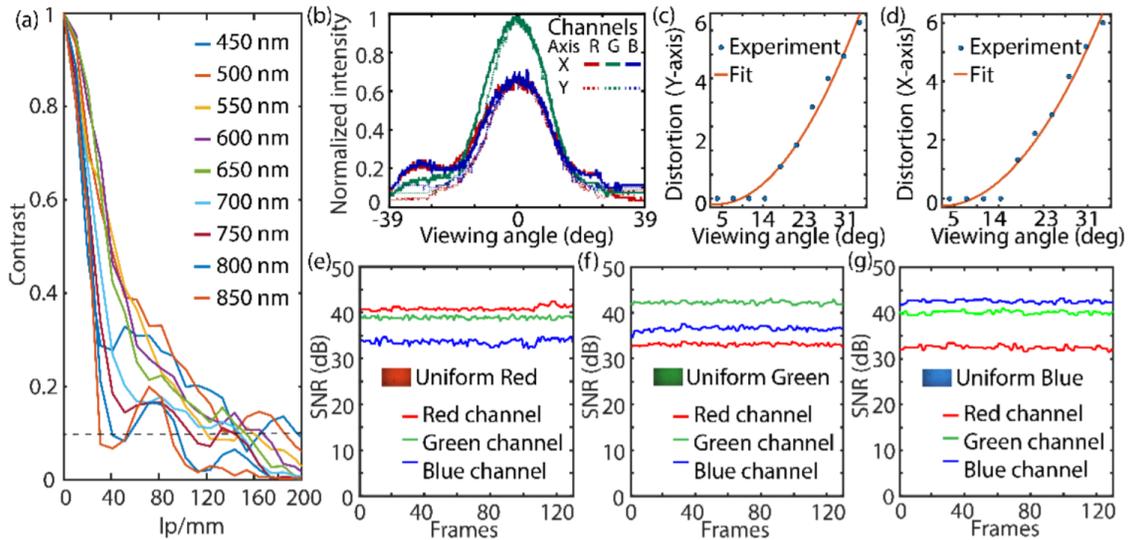

**Figure S8:** Aberration analysis of the low NA MDL. (a) Modulation-transfer-function (MTF) and (b) Vignetting measurement showing normalized intensity vs viewing angle in degrees. (c-d) Geometric distortion showing distortion of a regular geometric grid in the Y (c) and X (d) axes as a function of the viewing angle in degrees and (e-g) Signal-to-Noise Ratio (SNR) of the MDL when imaging a patch of uniform (e) Red, (f) Green and (g) Blue color for each of the Red, Green and Blue channels of the CMOS sensor.